\documentclass[preprint, 12pt, authoryear]{elsarticle}
\usepackage{amsmath, amsthm, mathtools}
\usepackage{amssymb, graphics, graphicx, color}

\usepackage{rotating}

\usepackage{algorithm}
\usepackage{algorithmic}

\newcommand{\bX}{{\bf X}}
\newcommand{\bA}{{\bf A}}
\newcommand{\bB}{{\bf B}}
\newcommand{\bU}{{\bf U}}
\newcommand{\bV}{{\bf V}}
\newcommand{\bS}{{\bf S}}

\newcommand{\bW}{{\bf W}}
\newcommand{\bH}{{\bf H}}
\newcommand{\bM}{{\bf M}}
\newcommand{\cA}{{\mathcal A}}
\newcommand{\cB}{{\mathcal B}}

\newtheorem{definition}{Definition}
\newtheorem{proposition}{Proposition}

\begin{document}
\begin{frontmatter}

\title{Nested Nonnegative Cone Analysis}

\author{Lingsong Zhang\corref{cor1}\fnref{fn1}}
\ead{lingsong@purdue.edu}
\address{Department of Statistics, Purdue University, 150 N. University St., West Lafayette, IN, 47907.}
\author{Shu Lu\fnref{fn2}}
\ead{shulu@email.unc.edu}
\address{Department of Statistics and Operations Research, University of North Carolina at Chapel Hill, CB \#3260, Chapel Hill, NC, 27599-3260.}
\author{J. S. Marron\fnref{fn3}}
\ead{marron@email.unc.edu}
\address{Department of Statistics and Operations Research, University of North Carolina at Chapel Hill, CB \#3260, Chapel Hill, NC, 27599-3260.}

\cortext[cor1]{Corresponding author}
\fntext[fn1]{Lingsong Zhang is Assistant Professor of Statistics in the Department of Statistics at Purdue University. }
\fntext[fn2]{Shu Lu is Assistant Professor of Operations Research in the Department of Statistics and Operations Research at the University of North Carolina at Chapel Hill.}
\fntext[fn3]{J. S. Marron is Amos Hawley Distinguished Professor of Statistics in the Department of Statistics at the University of North Carolina at Chapel Hill. }


\begin{abstract}
Motivated by the analysis of nonnegative data objects, a novel Nested Nonnegative Cone Analysis (NNCA) approach is proposed to overcome some drawbacks of existing methods. The application of traditional PCA/SVD method to nonnegative data often cause the approximation matrix leave the nonnegative cone, which leads to non-interpretable and sometimes nonsensical results. The nonnegative matrix factorization (NMF) approach overcomes this issue, however the NMF approximation matrices suffer several drawbacks: 1) the factorization may not be unique, 2) the resulting approximation matrix at a specific rank may not be unique, and 3) the subspaces spanned by the approximation matrices at different ranks may not be nested. These drawbacks will cause troubles in determining the number of components and in multi-scale (in ranks) interpretability. The NNCA approach proposed in this paper naturally generates a nested structure, and is shown to be unique at each rank. Simulations are used in this paper to illustrate the drawbacks of the traditional methods, and the usefulness of the NNCA method.
\end{abstract}
\end{frontmatter}
\section{Introduction \label{sec:intro}}
The advances of science and technology come along with large amounts of data. Many such contexts have complicated data structure, such as population of trees, data lying on manifolds, etc. Traditional statistical concepts and methods, are developed in the framework of (linear) Euclidean space without boundary constraints. Examples include mean and variance in introductory statistics, principal component analysis in multivariate statistics and in even more recent functional data analysis (FDA, \cite{ramsay2005functional, ferraty2006nonparametric}), These methods usually enjoy convenient properties such as linearity, since data are in an Euclidean space. However, the emerging complicated data objects such as trees, shapes and data that naturally lie in non-Euclidean space bring challenges to statistical analysis. One of the major challenges is that those objects may not lie in a linear space, and/or the data objects are restricted by a set of (mathematical or geometric) constraints. A new statistical framework, {\em object-oriented data analysis} (OODA), first named in \citet{wang2007object}, provides early systematic attempts at the analysis of complicated data objects.

\subsection{Nonnegative data and existing methods}
In this paper, a special class of object-data will be investigated, nonnegative data objects. Important examples include a number of types of spectral data (see e.g. \cite{marron2004time, li2005seldi}), and some tree structured objects (see e.g. \cite{wang2007object, shen2013functional}). Our main goal is to study the major modes of variation within a sample of nonnegative data objects. Principal Component Analysis (PCA) has been a central tool in multivariate analysis and FDA to reveal major modes of variation. The projections of a data set onto a set of orthogonal principal component (PC) directions, sequentially provides a decomposition of data into a series of orthogonal components with decreasing energy (i.e., decreasing eigenvalues). The sequence of subspaces spanned by these components are naturally nested within each other, i.e., adding additional PC components learn additional information than was available in the earlier components. However, PCA can be quite inappropriate for nonnegative data objects, because that some of the projections sometimes can easily leave the nonnegative orthant. Thus, these projections may not have physical interpretation, or the interpretation may not be sensible under some contexts. \cite{zhang2007singular} explored the impact of different types of centering on PCA, or more precisely on Singular Value Decomposition (SVD, a non-mean centered variation of PCA). Uncentered SVD has also been used to find important modes of variations, e.g. \citet{shen2005analysis, zhang2007singular}.  However, this origin centric method is even less robust against non-negativity, because in general the first direction will point into the interior of the orthant, so all orthogonal directions must point outside. An interesting matrix decomposition technique, Nonnegative Matrix Factorization (NMF), has been developed to overcome the problems of PCA/SVD, see e.g. \citet{paatero1994positive, lee1999learning}. Although it gains popularity in many machine learning applications, the NMF method suffers several severe drawbacks:

\begin{figure}
\begin{center}
\vspace{-0.2in}
\begin{tabular}{cc}
\includegraphics[angle=270, width=3in]{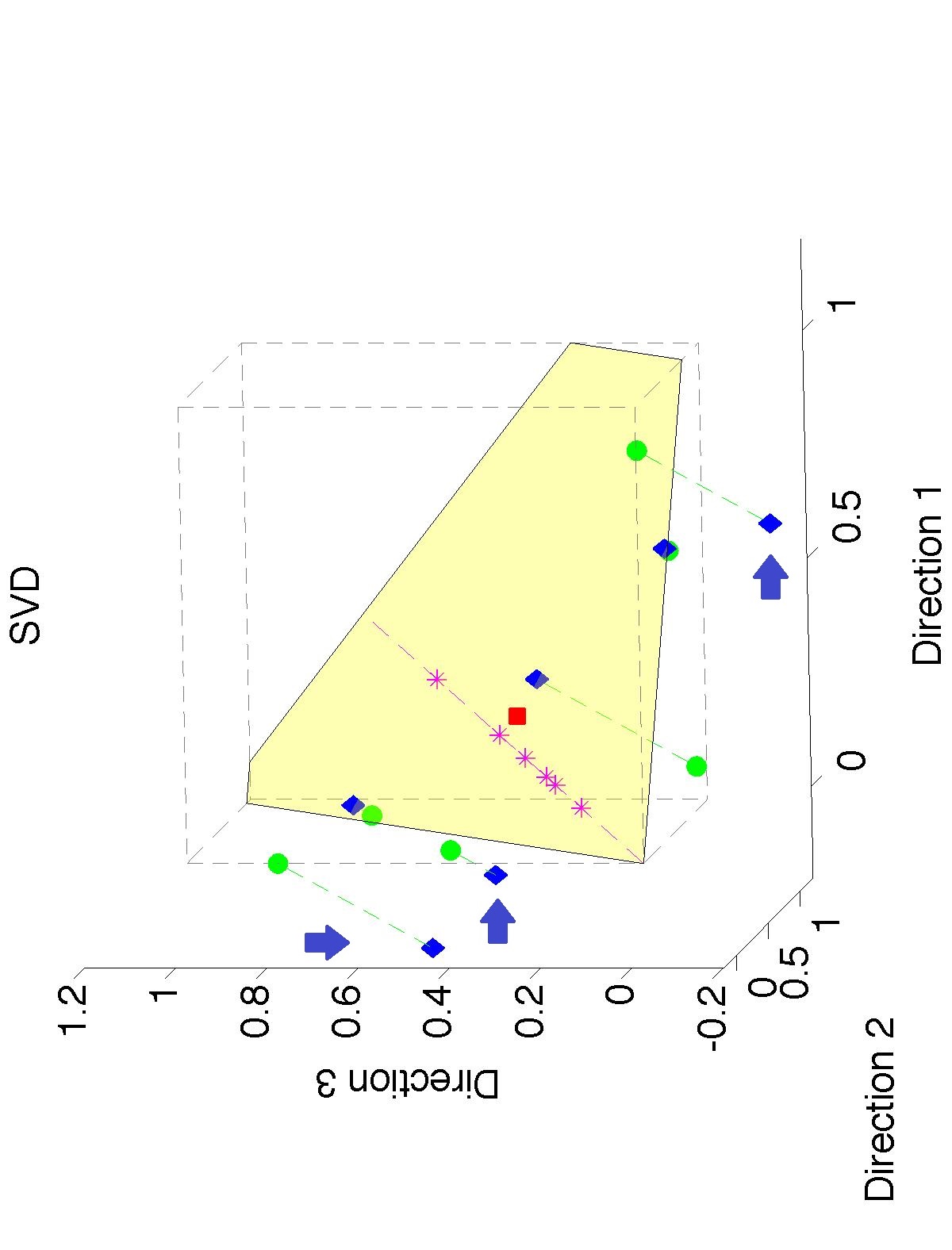} &
\includegraphics[angle=270, width=3in]{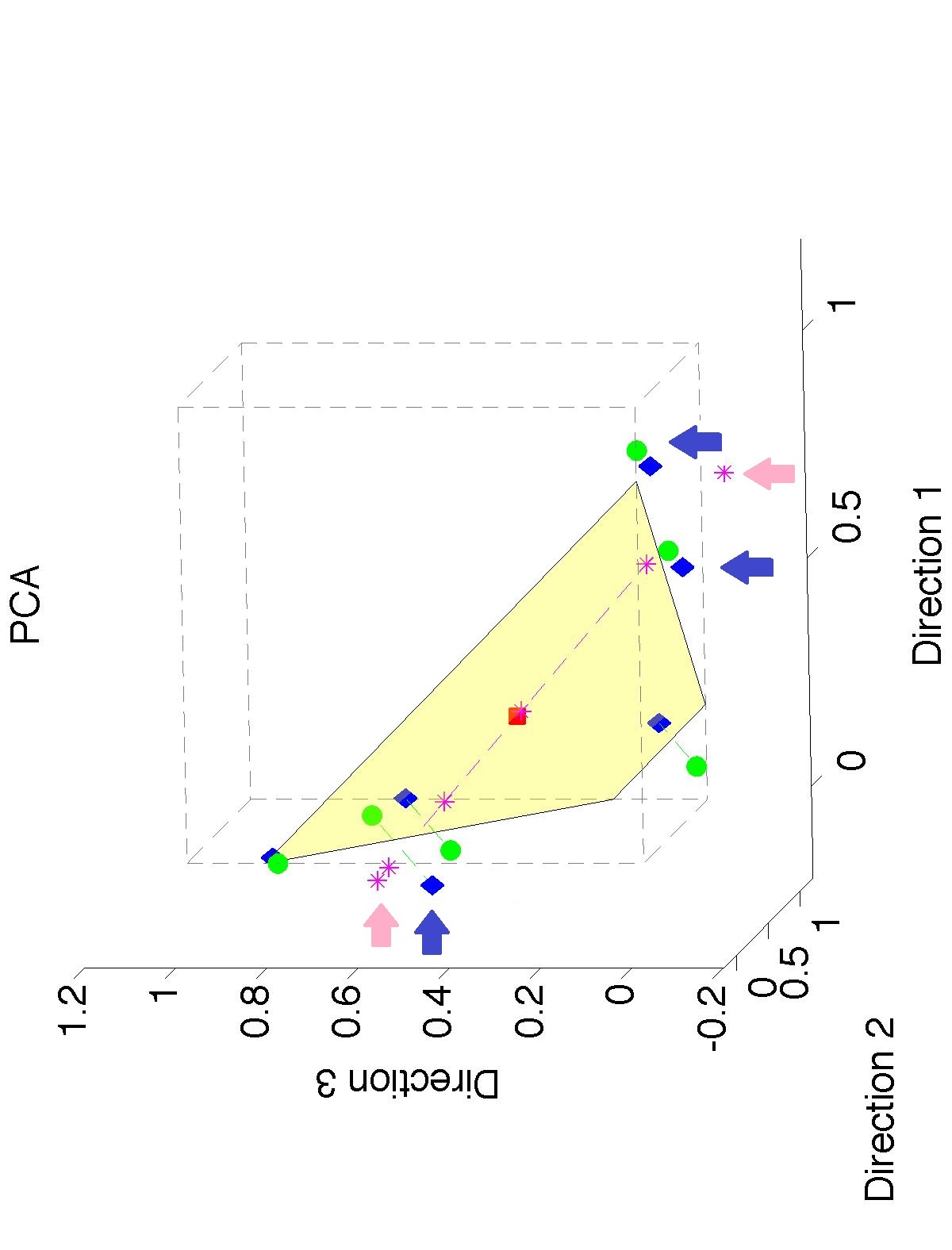} \\
\vspace{-0.1in}
\includegraphics[angle=270, width=3in]{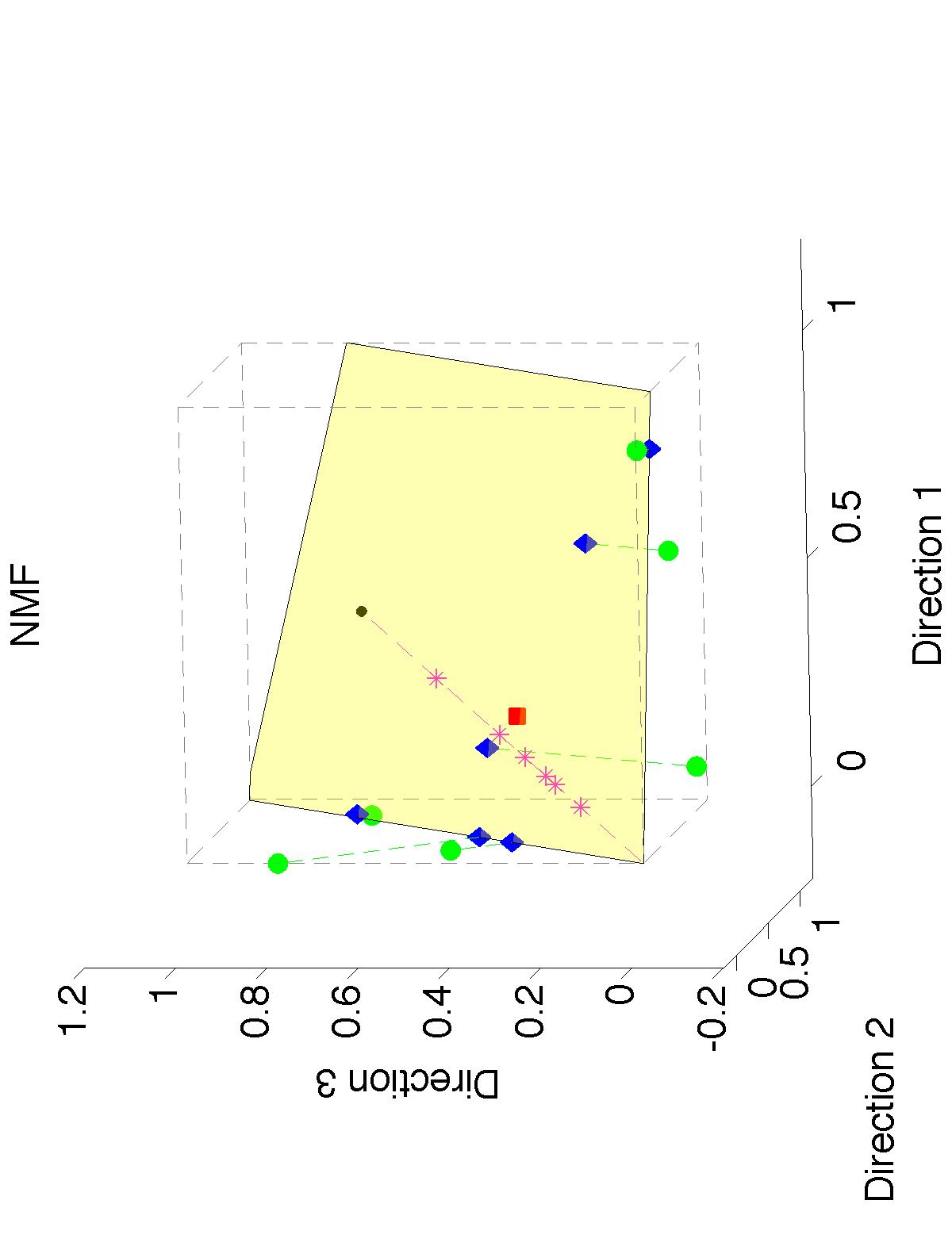} &
\includegraphics[angle=270, width=3in]{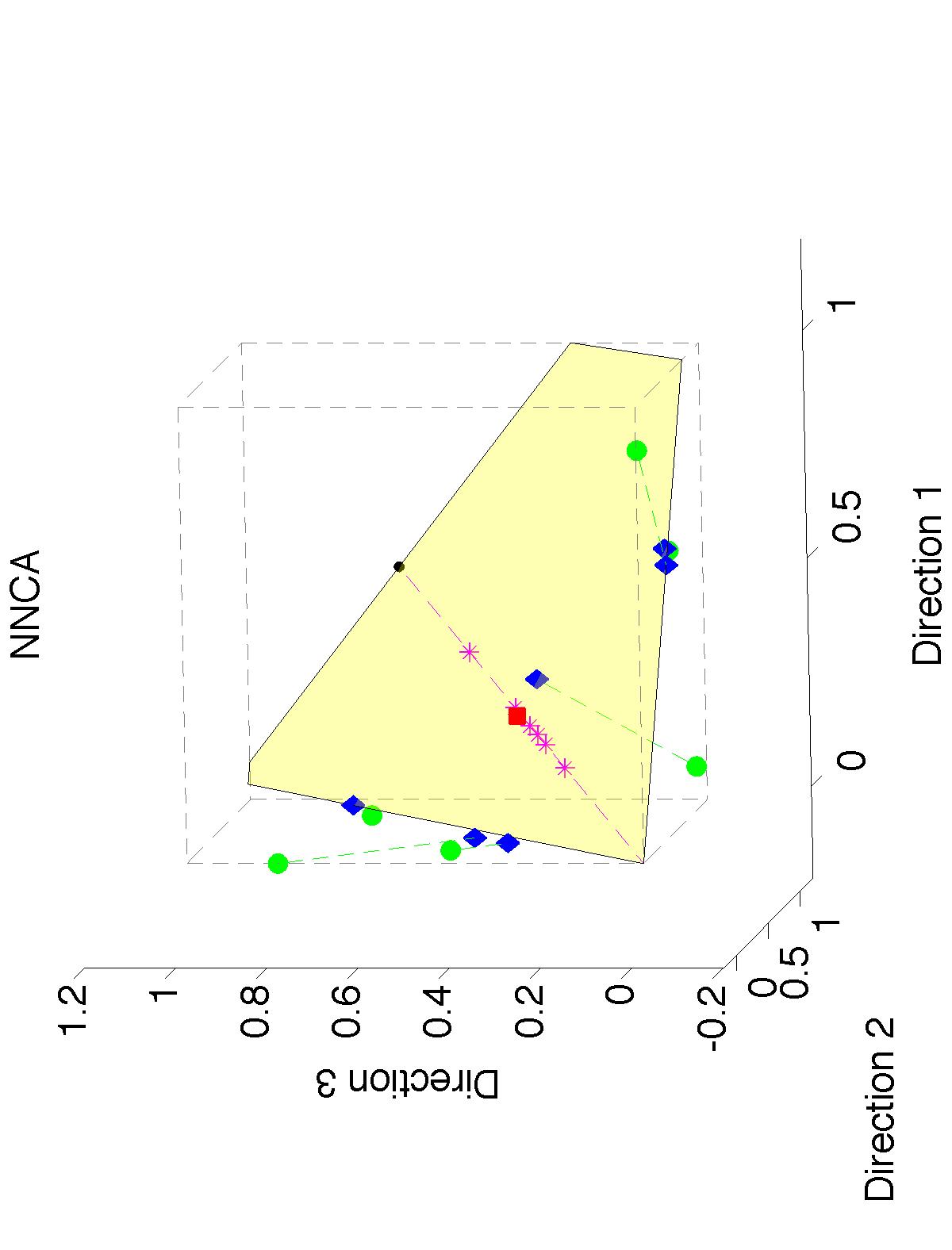} \\
\vspace{-0.1in}
\includegraphics[angle=270, width=3in]{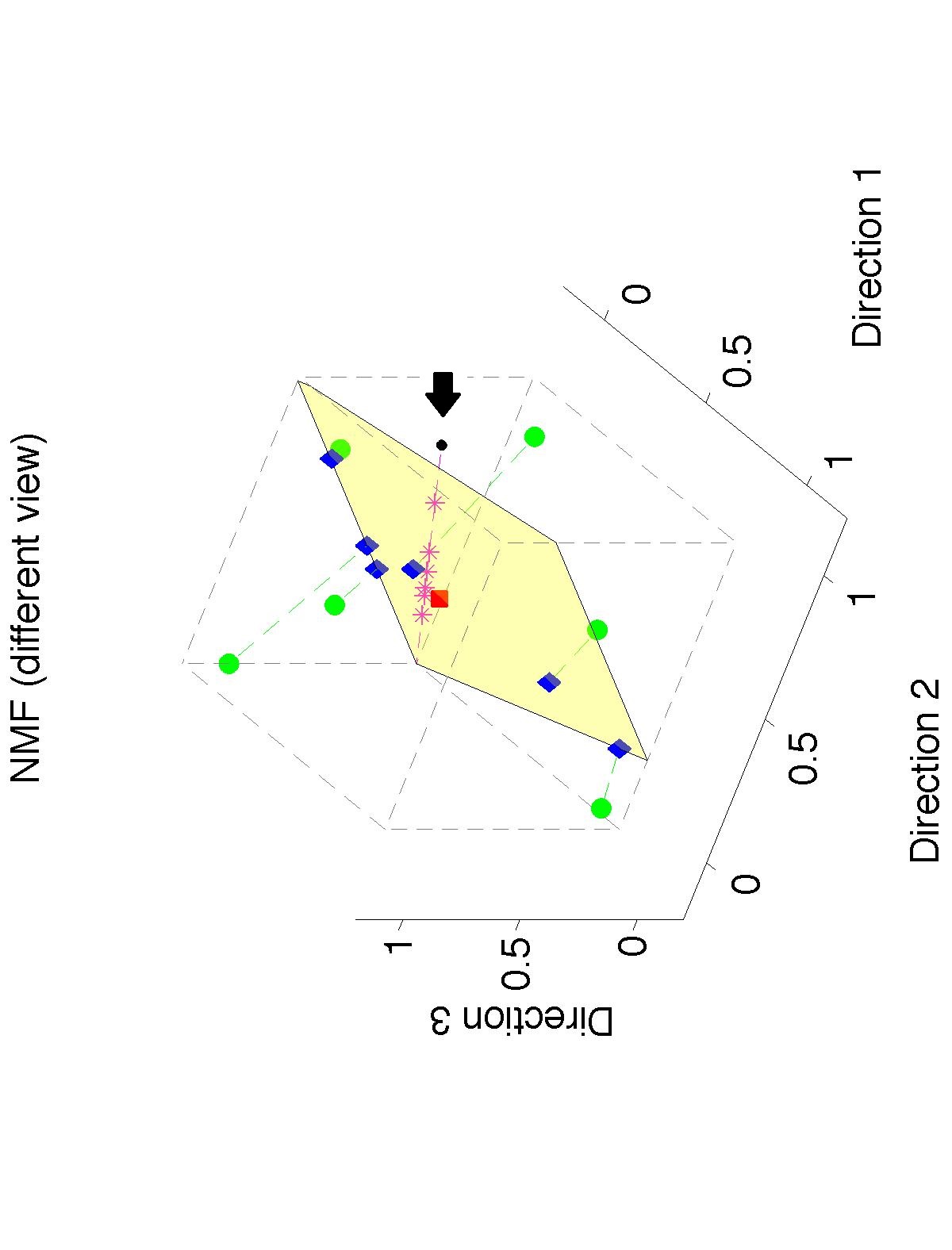} &
\includegraphics[angle=270, width=3in]{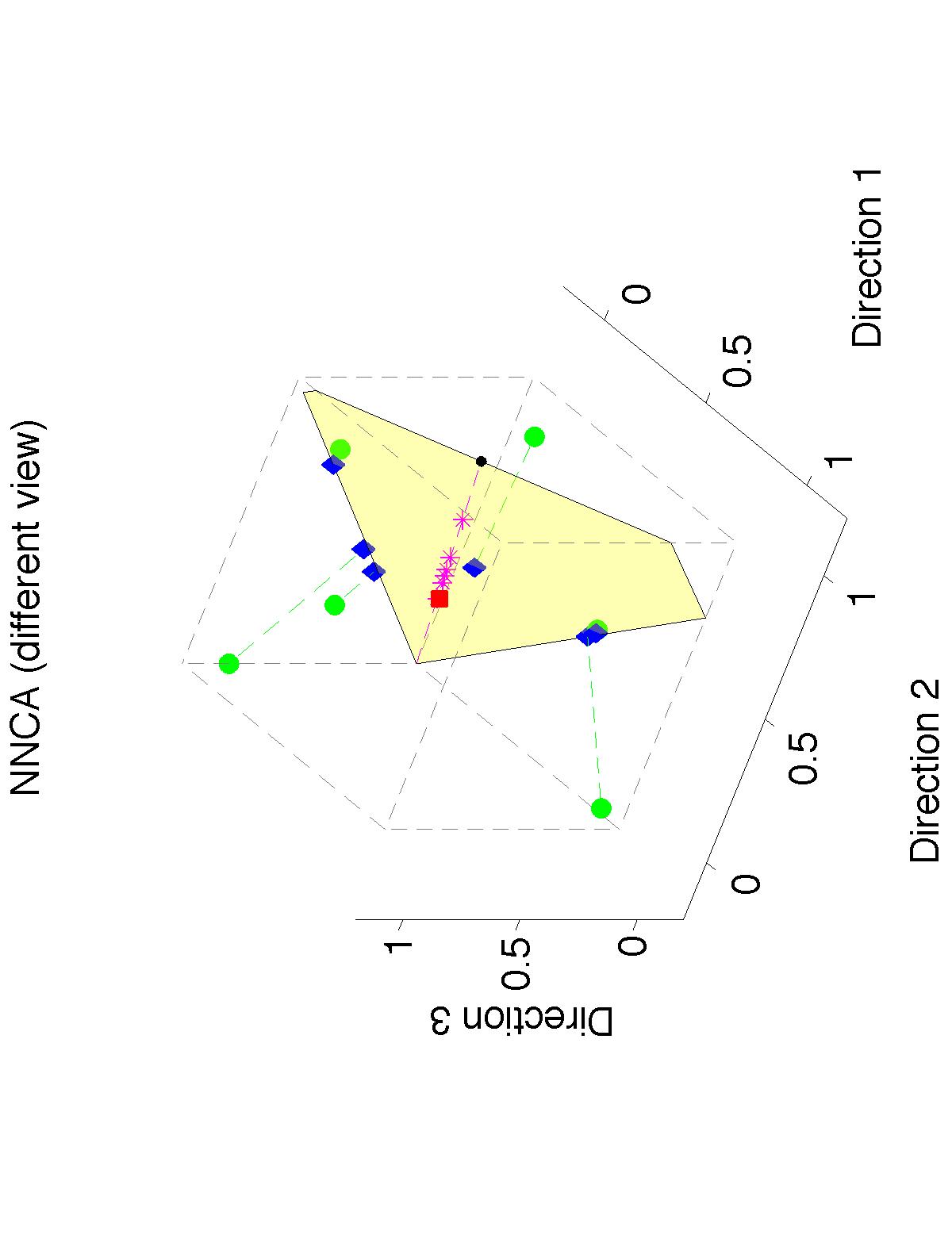} \\
\end{tabular}
\end{center}
\vspace{-0.2in}
\caption{\small \it The SVD, PCA, NMF and NNCA for a toy data set. The rank 1 subspaces (or affine space for PCA) are shown as the purple dashed lines, and the rank 2 subspaces are shown as the yellow planes. A unit box is plotted for better illustration.  The green dots are the original data points (same in all cases), the blue diamonds are their rank 2 approximations (several outside the orthant for PCA and SVD, as highlighted by the blue arrows), and the purple stars are their rank 1 approximations (some still outside for PCA as highlighted by the purple arrows). It shows that both PCA and SVD have approximations outside of the nonnegative cone. The bottom row is just a rotation of the middle row, which highlights the non-nested behavior of NMF. The subspaces of NMF at different ranks clearly are not nested within each other (as highlighted by the black arrow). The NNCA approach overcomes all these challenges.} \label{appmatfig}
\end{figure}

\begin{enumerate}
\item For a given rank $k$ approximation matrix, the NMF decomposition of this matrix may not be unique \citep{donoho2004does}. 
\item For a given rank $k$, the NMF approximation matrix may not be unique, as illustrated in Section \ref{subsec:onerun}. 
\item The space spanned by the NMF approximation at rank $k$ may not be nested in the space spaned by NMF aproximations at rank $j$, if $k\leq j$ as shown in the following toy example.
\end{enumerate}

\subsection{Drawbacks of PCA/SVD and NMF illustrated by a toy example}

A toy example is used to illustrate the drawbacks of PCA, SVD and NMF for the analysis of nonnegative data objects.  Each panel of Figure \ref{appmatfig} shows the same 6 observations (green dots) in the three dimensional nonnegative octant. This is one realization from a simulation setting that will be discussed in Section \ref{subsec:onerun}. The blue diamonds are the rank 2 approximations for each method, and the purple stars correspond to the rank 1 approximations. The projections from the original data to the rank 2 approximations are visualized by green dashed lines. The rank 2 approximating cones are shown as yellow planes (represented as quadrilaterals defined by their intersection with the graphical bounding box), and the rank 1 approximating cones are shown as purple lines (again intersected with the bounding box). The component-wise mean is highlighted as the red square in the plots. The SVD rank 2 and 1 approximating cones are the intersections of subspaces, a plane and a line through the origin, with the octant (top left panel).  The PCA rank 2 and 1 approximating sets are not actually cones, but instead are intersections of a plane and line not through the origin with the octant (top right panel). The NMF approximating cones are again intersections of a plane and line thorough the origin with the octant (middle and bottom left panels).  Both the SVD and PCA examples have three (ranking 2 approximating) points, highlighted with blue arrows that project to the approximating plane {\em outside} of the octant. 
The rank 1 PCA projections similarly have 3 points outside of the octant (highlighted with the purple arrows). All of the NMF projections lie within the octant, because NMF uses a sensible notion of projection of data to the approximating set, as shown in the middle and bottom left panels. The plots in the bottom row show only a rotated viewpoint, which is chosen to highlight the nesting aspects, while the viewpoint  in the middle row provided direct contrast with SVD and PCA. However this improved approximation of NMF comes at the price that the NMF rank 1 approximating cone is not nested in the rank 2 cone. The bottom left panel shows the intersection of the purple line (NMF rank 1 approximations) with one face of the unit box as a black dot (highlighted by the black arrow). This clearly reveals that the upper end of the purple line lies outside of the yellow quadrilateral. In this 3 dimensional example the angle between the plane and the line is 4.96 degrees. We conjecture that in higher dimensions this angle can be far bigger. The method proposed in this paper, the nested nonnegative cone analysis method, simultaneously yields non-negative and nested approximations, as shown in the middle and bottom right panels in Figure \ref{appmatfig}.

\subsection{From PCA/NMF to Nested Nonnegative Cone Analysis}
The goal of overcoming these problems of NMF by finding a suitably nested analog of PCA \citep{jung2010generalized}, motivates us to propose a novel {\em Nested Nonnegative Cone Analysis} (NNCA) approach to reveal underlying modes of variations within the data. The actual implementation uses the Backwards PCA idea of \cite{damon2013why}. PCA can be viewed as providing a sequence of nested affine spaces (indexed by rank of the approximation), where the projections of the data into these subspaces have the largest modes of variations for that rank. These nested subspaces usually are not suitable for the analysis of nonnegative data, because the orthogonality can cause the projections to easily leave the first orthant. However, this nested structure motivated us to identify a sequence of nested cones, where the projections of the data have the largest modes of variations for that rank. Note that this series of cones will generally be different from the usual PCA or SVD perpendicular sequences. The projections themselves will also form a sequence of approximations to the data of interest. Let $X_1$, $X_2$, $\cdots$, $X_n \in \mathbb{R}_+^d$ be $n$ $d$-dimensional observations that lie in the first orthant ($\mathbb{R}_+=\{x: x\geq 0\}$), i.e., the nonnegative cone. Let $K=\min(n, d)$. It is straightforward to know that these $n$ observations lie in a $K$-dimensional cone, which is also a sub-cone within the nonnegative cone. Define the data matrix $\bX=(X_1, \cdots, X_n)$. The NNCA approach is to identify a sequence of approximation matrices of $\bX$, $\{\bA_k\}$ ($k=1, \cdots, K$) indexed by rank, in the sense of minimizing the Frobenius norm of the residuals: $\min \|\bX-\bA_k\|_F^2$, where $\|\cdot\|_F$ is the Frobenius norm. Thus, an ideal sequence $\{\bA_k\}$ will satisfy
\begin{enumerate}
\item (Approximation optimality) $\bA_k$ is the best rank $k$ approximation to $\bX$, and also columns of $\bA_k$ are in $\mathbb{R}_+^d$, i.e., the $\{\bA_k\}$ are all nonnegative matrices.
\item (Nested Structure) For any $1\leq i\leq j \leq K$, the space spanned by the columns of $\bA_i$, is a subspace of the column space of $\bA_j$.
\end{enumerate}

The remaining part of this paper is organized as follows: Section \ref{sec:method} will introduce the NNCA method. The algorithms are discussed in Section \ref{sec:algorithm}. Some properties of the NNCA method will be discussed in Section \ref{sec:theory}. Section \ref{sec:simulation} provides more examples. Summary and further discussion will be in Section \ref{sec:discuss}.



\section{Nested nonnegative cone analysis \label{sec:method}}
In this section, we will introduce the nested nonnegative cone analysis approach. Let $\bX$ be the data matrix to be analyzed. The dimension of the matrix is $d\times n$, where all elements in it are nonnegative. In this paper, the inequality $\bX \geq 0$ is used to denote that $\bX$ is a nonnegative matrix, as is common in the field of optimization. Furthermore, let $r_0=\mathrm{rank}(\bX)$. Usually $r_0=\min(d, n)$. We will first review the notation and mathematics behind PCA/SVD and NMF in Section \ref{SVDNMFsec}, and then introduce the NNCA method in Section \ref{NNCAintrosec}. 
\subsection{PCA/SVD and NMF \label{SVDNMFsec}}
The SVD of $\bX$ is defined as 
\begin{equation}
\bX=\bU\bS\bV^T, \label{svd-original-definition}
\end{equation}
where $\bU$ is a $d\times r_0$ matrix, $\bS$ is an  $r_0\times r_0$ diagonal matrix, and $\bV$ is an $n\times r_0$ matrix. In addition,  $\bU^T\bU=\bV^T\bV=I_{r_0}$, and the diagonal elements in $\bS$ form a non-increasing sequence: $s_1\geq s_2\geq \cdots 
\geq s_{r_0}$. Let $u_i$ be the $i$th column in $\bU$, $v_j$ be the $j$th column in $\bV$, and $\bA_k=\sum_{i=1}^k s_i u_iv_i^T$, where $k\leq r_0$. It can be shown that $\bA_k$ provides the best rank $k$ approximation to $\bX$ \citep{eckart1936approximation}, in the sense that
\[
\bA_k=\mathop{\mathrm{argmin}}_{\{\bA: \mathrm{rank}(\bA)=k\}} \|\bX-\bA\|_F^2.
\]
In this paper, each rank 1 matrix $s_i u_iv_i^T$ is called an {\em SVD component}.

The PCA of $\bX$ is defined as the eigen-decomposition of the covariance matrix of $\bX$. Let $m=\frac{1}{n}\bX{\bf 1}_{n\times 1}$ be the sample mean vector of $\bX$, i.e., this is the mean of the columns in $\bX$.  The sample covariance matrix can be calculated as
\[
\frac{1}{n-1}(\bX-\bM_c)(\bX-\bM_c)^T,
\]
where $\bM_c= m{\bf 1}_{1\times n}$ is called the column mean matrix. It can be shown that the eigenvectors of the sample covariance matrix can be calculated as the singular value decomposition of $\bX-\bM_c$. Let
\begin{equation}
\bX-\bM_c=\bU_c\bS_c\bV_c^T. \label{pca-original-definition}
\end{equation}
Here the columns in $\bU_c$ can be shown to be the principal component directions, and $\bV_c\bS_c$ are the principal component loadings. Let $\bA^c_k=\sum_{k=1}^k s^c_iu^c_iv^c_i$, where $u^c_i$, $v^c_i$ is the $i$th column in $\bU_c, \bV_c$ respectively, and $s_i^c$ is the $i$th largest singular value in $\bS_c$. Then
\[
\bA^c_k=\mathop{\mathrm{argmin}}_{\{\bA: \mathrm{rank}(\bA)=k\}} \|(\bX-\bM_c)-\bA\|_F^2.
\]

As discussed above, the SVD/PCA provides the best rank $k$ approximation to $\bX$/$(\bX-\bM_c)$ respectively. In addition, SVD/PCA can be used sequentially identify a series of approximation matrices at different ranks ($k$),  in either a forwards or a backwards fashion. The forward fashion means that we start from $k=1$, and then increase the rank to 2, $\cdots$, until $k=r_0$. In the backward approach we can start from $k=r_0$, and then decrease the rank to $r_0-1$, $\cdots$, down to 1. It can be shown that for Euclidean data, the forward SVD/PCA is equivalent to the backward SVD/PCA method, using standard analysis of variance decompositions of sums of squares. However, for non-Euclidean data, these two approaches may not be the same, as illustrated in \cite{jung2010generalized}. In the following sections, we simplify the above triplet-factorization in equations \eqref{svd-original-definition} and \eqref{pca-original-definition}. Take the equation \eqref{svd-original-definition} as an example, the SVD is recast as
\[
\bX=\bU^*(\bV^*)^T,
\]
where $\bU^*$ is the same as the $\bU$ in equation \eqref{svd-original-definition}, and $\bV^*=\bV\bS$. Note that essentially, we assume that the columns in $\bU^*$ have unit $L_2$ norms. For easy presentation, we will skip the asterisk from now on.

From the above definitions, columns in all these $\bU$ matrices are orthogonal to each other, and thus the corresponding low rank approximation (when $k \geq 2$) can take on negative values, even though $\bX$ is nonnegative. This led to the invention of NMF, see early developments of NMF in \cite{paatero1994positive} and \cite{lee1999learning}. See \cite{berry2007algorithms} for a review of different NMF algorithms. One type of NMF of $\bX$ is defined as follows:
\begin{equation}
(\bW, \bH)=\mathop{\mathrm{argmin}}_{\bW \geq 0, \bH \geq 0} \|\bX-\bW\bH^T\|_F^2, \label{nmfdefinition}
\end{equation}
where $\bW$ is a $d\times k$ matrix, $\bH$ is a $n\times k$ matrix, and $k\leq r_0$. In addition, the columns of $\bW$ and $\bH$ are linearly independent, and columns of $\bW$ have unit $L_2$ norms. However, the NMF suffers several major drawbacks listed below:
\begin{enumerate}
\item The factorization pair $(\bW, \bH)$ is not unique; 
\item The approximation matrix $\bW\bH^T$ is not unique;
\item The space spanned by the columns in $\bW$ at rank $i$, may not be a subspace of the space spanned by the columns of $\bW$ at rank $j$ ($i\leq j$), i.e., the approximation subspaces are not nested within each other. 
\end{enumerate}
All of these drawbacks of earlier methods are summarized in Table \ref{ProsConsTable}, and further illustrated in Section \ref{sec:simulation}.  This provides motivation for the development on NNCA.

\begin{table}
\caption{Pros and Cons of PCA/SVD and NMF methods} \label{ProsConsTable}
\begin{tabular}{l|l|l} \hline\hline
 & Pros & Cons \\ \hline\hline
SVD & 1) globally best rank $k$ approximation & tends to leave nonnegative cone\\
       & 2) forward and backward SVD are the same & \\
       & 3) different ranks are nested within each other & \\ \hline
PCA & 1) globally best rank $k$ approximation  & tends to leave nonnegative cone\\
	&  \qquad after removing the mean & \\
       & 2) forward and backward PCA are the same & \\
       & 3) different ranks are nested within each other &\\
       & 4) when applicable, good interpretation & \\ 
       & \qquad  (largest variability) & \\ \hline
NMF & 1) locally optimal approximation & 1) factorization may not be unique \\
        & 2) always within nonnegative cone & 2) approximation matrix may not \\
        & & \qquad  be unique \\
        & & 3) no forward NMF\\
        & & 4) different ranks not be nested  \\  \hline \hline      
\end{tabular}
\end{table}

\subsection{Nested Nonnegative Cone Analysis\label{NNCAintrosec}}
Here we propose the novel Nested Nonnegative Cone Analysis (NNCA) approach. A forward approach to NNCA is especially challenging, because it has to ensure that the residuals, at each level of approximation, are nonnegative.  A direct formulation of the needed optimization would result in an extremely complicated problem.  \cite{damon2013why} noted similar phenomena in a variety of other learning contexts, and recommend a general backwards approach, called {\em Backwards PCA}, based on the idea of a "series of nested constraints". Thus, we apply instead a backward approach. We will first define the rank $r_0-1$ approximation, and then sequentially define rank $k$ approximation as $k$ decreases. 

The rank $r_0-1$ NNCA approximation matrix is defined as follows:
\begin{definition}
A rank $r_0-1$ NNCA approximation is defined as a solution of 
\begin{equation}
\min_{\bA}\|\bX-\bA\|_F^2, \textnormal{ subject to } \mathrm{rank}(\bA)=r_0-1, \textnormal{ and }\bA\geq0. \label{largestrankequation}
\end{equation} 
\end{definition}
As in Section 2.1, $\bA\geq0$ means that all elements in $\bA$ are nonnegative. To highlight the rank $r_0-1$ approximation, a solution to equation \eqref{largestrankequation} is denoted as $\bA_{(r_0-1)}$. Let $s_1 \geq \cdots \geq s_r$ be the decreasing sequence of singular values of $\bX$. If $s_{r_0-1} >s_{r_0}$, we conjecture that this solution will be unique.


If we know a rank $k+1$ approximation matrix $\bA_{k+1}$ to $\bX$, a rank $k$ approximation $\bA_k$ to $\bX$ is then defined as follows:

\begin{definition}
Let $\bA_{k+1}$ be a rank $k+1$ ($k=1, \cdots, r_0-2$) NNCA approximation to $\bX$, a rank $k$ approximation, $\bA_k$, is the solution of 
\begin{equation}
\min_{\bA} \|\bA_{k+1}-\bA\|_F^2, \textnormal{ subject to } \mathrm{rank}(\bA)=k, \textnormal{ and } \bA\geq0. \label{lowerrankequation}
\end{equation}
\end{definition}

Thus, $\bA_{r_0-1}$, $\bA_{r_0-2}$, $\cdots$, $\bA_1$ is a sequence of approximation matrices of $\bX$, indexed by rank. This sequence $\{\bA_i\}$ ($i=r_0-1, r_0-2, \cdots, 1$) is called the {\em NNCA approximating sequence}. If $s_{k}>s_{k+1}$, where $s$ are defined earlier, we also conjecture that $\bA_k$ is unique.

Note that if the nonnegativity constraints are not imposed, the above NNCA approximation sequence corresponds to a backward SVD sequence. In addition, if all different $\bA_{k+1}$ in Definition 2 are replaced by $\bX$, then this sequence of NNCA is then equivalent to an NMF sequence using the least square loss function. However, as discussed earlier, the NMF sequence is generally not nested. Our definition imposes a very natural nested structure. 

\section{The NNCA algorithm \label{sec:algorithm}}

The core part of the NNCA approach is the equations \eqref{largestrankequation} and \eqref{lowerrankequation}. Even though these two problems seem to have simple mathematical forms, the searching of the optimal $\bA$ in each equation is still quite challenging. One of the reasons is that projections onto a cone are not all standard Euclidean projections onto a subspace. Note that the two optimization problems share a common optimization structure that can be written as follows: let $\bB \geq 0$ have rank $k+1$, our target is to find the solution of
\begin{equation}
\min_{\bA} \|\bB-\bA\|_F^2, \textnormal{ subject to } \mathrm{rank}(\bA)=k, \textnormal{ and } \bA \geq 0. \label{nncaproblem0}
\end{equation}
Note that the feasible set defined by the constraint $\mathrm{rank}(\bA)=k$ is not a closed set. And thus, we reformulate the problem to be
\begin{equation}
\min_{\bA} \|\bB-\bA\|_F^2, \textnormal{ subject to } \mathrm{rank}(\bA)\leq k, \textnormal{ and } \bA \geq 0. \label{nncaproblem}
\end{equation}
The optimal solution of \eqref{nncaproblem} often turns out to be of rank $k$. In those cases, \eqref{nncaproblem} is equivalent to \eqref{nncaproblem0} in the sense that they have the same optimal solutions. In any case, \eqref{nncaproblem0} and \eqref{nncaproblem} always have the same optimal value, because any matrix of rank less than $k$ can be changed into a slightly different matrix of rank k. The constraint  $\mathrm{rank}(\bA) \leq k$ corresponds to a closed set, and is easier to handle analytically.

A major challenge in \eqref{nncaproblem} is the combination of $\bA\geq0$ and $\mathrm{rank}(\bA)\leq k$. Because of the rank constraint, the above problem is a nonconvex optimization problem. In general, nonconvex optimization problems are very hard to solve, see, e.g., \cite{murty1987some, vandenberghe1996semidefinite, ben2001lectures, boyd2004convex}). In this subsection, we provide an approximating algorithm which is based on SVD. Note that the rank$(\bA)\leq k$ constraint can be reformulated to $\bA=\bU\bV^T$, where $\bU$ is a $d\times k$ matrix, and $\bV$ is $n\times k$ matrix. Thus, the above problem can also be written as
\[
\min_{(\bU, \bV)} \|\bB-\bU\bV^T\|_F^2, \textnormal{ subject to }  \bU\bV^T \geq 0.
\]
Note that for identifiability, we may want to assume columns in $\bU$ all have unit $L_2$ norms. It is straightforward to see that this is not a convex problem over the pair $(\bU, \bV)$. However, conditional on $\bU$ (or $\bV$), this problem can be viewed as convex optimization problems over $\bV$ (or $\bU$) respectively.

Note that the above reformulation is similar to both the SVD and NMF formulations. The SVD approach does not have the constraint $\bA\geq0$ (i.e., $\bU\bV^T\geq 0$), while the NMF uses a stronger constraint that both $\bU$ and $\bV$ are nonnegative. Due to the non-uniqueness of many NMF algorithms, we will develop our NNCA based on the SVD approximation, as shown in the next subsection.

\subsection{SVD-based NNCA}
A central algorithm in the nested nonnegative cone analysis approach is to identify the best rank $k$ nonnegative approximation matrix $\bA_s$ for the rank $k+1$ matrix $\bB$. Recall that the matrix $\bB$ has the dimension $d\times n$.  In this subsection, we will introduce an SVD-based algorithm.

The target of our algorithm is identification of low rank or low dimensional approximations of the original data. Note that without non negativity constraints, SVD can be used to identify the best rank $k$ approximation $\bA_s$ for $\bB$. If $\bA_s$ is a nonnegative matrix, then $\bA_s$ is the $\bA$ matrix we plan to find. Otherwise, let $(\bU_s, \bV_s)$ be the corresponding SVD decomposition. It can be shown that the span of $\bU_s$ is the best $k$-dimensional approximating subspace, which has the largest variability (sum of squares) among all projections of $\bB$ to such $k$-dimensional subspaces (see Proposition \ref{optimalsubspacetheorem}). The $\bV_s$ are the corresponding coefficients of these projections. Note that, under this context, these projections are out of the nonnegative cone. We will then identify the convex set $\mathcal{C}_s=\mathrm{span}(\bU_s)\cap \mathbb{R}_+^d$, and project $\bB$ onto $\mathcal{C}_s$. The projections of $\bB$ onto $\mathcal{C}_s$ form the resulting $\bA$. This SVD-based NNCA algorithm is summarized in Algorithm \ref{algorithm:SVD-NNCA}. Note that the (A1) step can be viewed as a modified version of a nonnegative least squares problem \citep{lawson1974solving}, so standard optimization routines can be applied to solve it.


\begin{algorithm}[ht]
\caption{SVD-NNCA }\label{algorithm:SVD-NNCA}
Core algorithm of the SVD-based Nested Nonnegative Cone Analysis method: identify an rank $k$ nonnegative approximation matrix $\bA$ to a rank $k+1$ nonnegative matrix $\bB$. 
\hrule height .1pt
\begin{algorithmic}[1]
\STATE Identify the best rank $k$ approximation matrix $\bA_s=\bU_s\bV_s^T$ by a standard SVD of $\bB$.
\IF{$\bA_s$ is nonnegative}
	\STATE $\bA \gets\bA_s$ and stop.
\ELSE	
	\STATE Minimize $\|\bB-\bU_s\bV^T\|_F^2$ over $\bV$, subject to $\bU_s\bV^T\geq0$. \hfill (A1)
	\STATE Let $\bV$ be the above minimizer of (A1-1), $\bA \gets \bU_s\bV$.
\ENDIF

\end{algorithmic}
\end{algorithm}

\section{Mathematical Properties of the NNCA method \label{sec:theory}}
From the above definition, it is straightforward to see that the $\mathrm{span}(\bA_k) \subset \mathrm{span}(\bA_{k+1})$ for all $k$. So the SVD-based NNCA approach naturally generates a nested structure. In this section, we further show some properties of the proposed NNCA algorithms. 

In this section, let $d(x, y)$ be the distance between two vectors $x$ and $y$, for example the Euclidean distance  $d(x, y)=(x^Ty)^{1/2}$. Further let $\mathcal{S}$ be a subspace, recall that the distance of $x$ to $\mathcal{S}$ is defined as the smallest distance between $x$ and any vector in $\mathcal{S}$, i.e., $d(x, \mathcal{S}) =\min \{d(x, y):  y \in \mathcal{S}\}$. Recall that $\bX$ is the $d\times n$ matrix formed by $n$ vectors, $x_1$, $\cdots$, $x_n$ (the columns of $\bX$). The square distance of $\bX$ to $\mathcal{S}$ in this section is defined to be the sum of squared distances of  the $x_i$ to $\mathcal{S}$, i.e., $d^2(\bX, \mathcal{S})=\sum_{i=1}^n d^2(x_i, \mathcal{S})$. 

The following proposition shows that the subspace spanned by $\bU_s$, the rank $k$ SVD of $\bB$ (here rank$(\bB)=k+1$), provides the best rank $k$ approximation to $\bB$, when there is no nonnegativity constraint.  Let $s_1 \geq s_2 \geq \cdots \geq s_k \geq s_{k+1}$ be the $k+1$ singular values of $\bB$.

\begin{proposition}
In the SVD-NNCA algorithm, if $s_k > s_{k+1}$, $\mathrm{span}(\bU_s)$ provides the best rank $k$ subspace approximating the columns of $\bB$, i.e., the $\mathrm{span}(\bU_s)$ is minimizer of the square distance
to $\bB$ over all rank $k$ subspaces. \label{optimalsubspacetheorem}
\end{proposition}
The proofs for Proposition 1 and Proposition 2 are provided in the appendix for better presentation.

The following proposition shows that this nonnegative approximation is unique, under a condition that is similar to the uniqueness of SVD.
\begin{proposition}
If $s_k > s_{k+1}$, the nonnegative rank $k$ approximation $\bA$ in (A1) is unique. \label{uniquenesstheorem}
\end{proposition}
Note that if $s_k > s_{k+1}$,  then $\bU_s\bV_s^T$ is unique, i.e., the $k+1$ SVD component ($s_{k+1}u_{k+1}v_{k+1}^T$) is identifiable. This leads to the uniqueness of $\bA_k$. This condition is mild, because it only requires that the $k$th component and $(k+1)$th component are distinguished from each other. This proposition states that the one-way SVD-NNCA method is also unique under this mild condition.  Note that it is about uniqueness of the approximating spaces. The $k$-th individual component is similarly unique when $s_{k-1} > s_k > s_{k+1}$. The case of non-distinguishable components can be theoretically analyzed by working with subspaces, but that is beyond the scope of the present paper.

It is worth mentioning that the optimization problems considered in NMF (equation \eqref{nmfdefinition} in Section 2.1), the stated goal of NNCA definition (equation \eqref{nncaproblem0} in Section 3) and the SVD-based NNCA algorithm (A1) are closely related. Let us focus on a rank $k+1$ input matrix $\bB$. A common goal of all three is the identification of a rank $k$ nonnegative approximating matrix $\bA$ to $\bB$, which is defined in the equation \eqref{nncaproblem0}. If the solution $\bA$ of equation \eqref{nncaproblem0} can be factorized as two nonnegative matrices (although such factorization may not generally exist, see e.g. \cite{berman1973rank, thomas1974rank}), the NMF algorithm (with sufficiently many random restarts)  can be used to identify such $\bA$.  Meanwhile, if we apply SVD algorithm to $\bB$, and the resulting SVD rank $k$ approximation matrix is nonnegative, our SVD-based NNCA algorithm then will provide the global optimal solution to the problem in equation \eqref{nncaproblem0}. In general, the resulting matrix $\bA$ from the SVD-based NNCA method is an approximating solution of the following problem, in which the squared distance $d^2(\bB, \mathrm{span}(\bA))$ is a penalty term with a large penalty parameter $\lambda>0$:
\[
\min \|\bB-\bA\|_F^2+\lambda d^2(\bB-\mathrm{span(\bA)}), \textnormal{subject to  rank}(\bA)\leq k, \textnormal{ and } \bA\geq 0.
\]

\section{Simulations \label{sec:simulation}}
In this section, we use some simulations to further illustrate the usefulness of NNCA methods over PCA/SVD and NMF methods. For better illustration, the toy examples in this section are in the 3-dimensional nonnegative cone. In Section \ref{subsec:onerun}, we will simulate one realization of data with sample size $n = 6$, i.e. a $3\times 6$ matrix, and will compare the performance of these four methods. In Section \ref{subset:repeatrun}, we will repeat the same simulation as in Section \ref{subsec:onerun} 100 times, and report summarized performance comparisons. All the elements in these matrices are independently simulated from $U[0, 1]$, except its $(3, 1), (3, 2), (3, 3), (1, 4), (1, 5), (1, 6)$ and $(2, 6)$ entries are zero. In Section \ref{subsec:nonnested}, we will further illustrate the non-nested issues of the NMF method by another simulation. 

\subsection{One simulation realization \label{subsec:onerun}}

\begin{table}
\caption{The cumulative approximation matrices (rounded to two decimal points) indexed by ranks for different methods. It shows that the PCA method easily leaves the nonnegative cone, and in fact 50\% of the projections can leave the octant.. The SVD method has the approximating matrix at rank 1 within the cone (Perron-Frobenius theorem), but three of the rank 2 approximation points leave the cone. The observations leaving the cone are highlighted using a bold font. NNCA solves the above non-negativity issues, and also has sparse results for the cumulative of the first two components. The NMF method has non-sparse nonnegative average approximating matrices for both ranks.} \label{appmattable}
\begin{center}
\begin{tabular}{c|c|c} \hline\hline
Method & Rank index & Approximation matrix\\ \hline\hline
PCA & 1 & $\begin{pmatrix*}[r]
    0.28    &    {\bf 0.82}    &     0.61   &     0.08   &    {\bf -0.10}  &    {\bf -0.07}   \\
   0.38 &{\bf 0.28} &0.32 &0.42 & {\bf 0.45} &{\bf 0.45} \\
   0.32 &  {\bf -0.15} &0.03 &0.49 & {\bf 0.68} & {\bf 0.62}
   \end{pmatrix*}$ \\ \cline{2-3}
   & 2 & $\begin{pmatrix*}[r]
    0.18  &      {\bf  0.87}  & {\bf 0.59} & 0.10  &  {\bf -0.14}  & 0.01 \\
    0.90 &  {\bf -0.00} &{\bf 0.45} &0.27 & {\bf 0.66} & 0.01 \\
    0.09 &  {\bf -0.03} &{\bf -0.03} &0.56 &{\bf  0.56} &  0.81
   \end{pmatrix*}$\\ \hline
SVD & 1 & $\begin{pmatrix}
    0.21   &     0.09   &    0.17   &     0.13   &    0.30   & 0.14  \\
    0.51  &  0.22  &  0.42  &  0.31  & 0.74  &0.35 \\
    0.38 & 0.17 & 0.31 &0.23 & 0.55 &   0.26
    \end{pmatrix}$ \\\cline{2-3}
    & 2 & $\begin{pmatrix*}[r]
    0.33   & {\bf 0.70}   & 0.62   &{\bf -0.07}   &0.03   &  {\bf -0.23} \\
    0.53  &  {\bf 0.28}  & 0.47  &{\bf 0.30}  & 0.71  & {\bf 0.31} \\
    0.30 & {\bf -0.25} & 0.01 & {\bf 0.37} & 0.73 & {\bf 0.51}
    \end{pmatrix*}$\\ \hline
NNCA & 1 & $\begin{pmatrix}
    0.27  &      0.22  & 0.24  &0.16  &0.36  & 0.20 \\
    0.53  &  0.43  &  0.46  & 0.32  & 0.71  &0.40\\
    0.35 &0.29 &0.31 &0.21 &0.47 &0.27
    \end{pmatrix}$ \\\cline{2-3}
    & 2& 
    $\begin{pmatrix*}[r]
        0.33 &   0.59   & 0.62   & 0   &0.03   & 0  \\
        0.53  &  0.44     &  0.47   & 0.32   &0.71   & 0.40 \\
        0.30 & 0 & 0.01 &  0.34 & 0.73 & 0.43
         \end{pmatrix*}$
    \\ \hline
NMF$^*$  & 1 & $\begin{pmatrix}
0.21&0.09&0.17&0.13&0.3&0.14 \\
0.51&0.22&0.42&0.31&0.73&0.35 \\
0.38&0.17&0.32&0.23&0.55&0.26
    \end{pmatrix}$\\\cline{2-3}
    & 2 &
    $\begin{pmatrix*}[r]
0.19&0.88&0.66&0&0.01&0 \\
0.44&0.21&0.31&0.33&0.72&0.41 \\
0.4&0&0.16&0.33&0.73&0.42
    \end{pmatrix*}$ \\ \hline\hline
\end{tabular}
\end{center}
{\footnotesize NMF$^*$ stands for the best approximating matrix (the one whose residual matrix has the smallest Frobenius norm) over the 100 different repeated runs.} 
\end{table}

In this subsection, we compare different methods (SVD, PCA, NMF and NNCA) by a specific simulation realization. The data set we simulated is (rounded to two decimal points)
\[
X=\begin{pmatrix*}[r]
    0.09  &  0.90  &  0.62     &    0   &      0      &   0 \\
    0.85  &  0.02  &  0.47    &0.20   & 0.75      &   0\\
         0  &      0    &     0   & 0.45   & 0.70  &  0.80
\end{pmatrix*},
\]
with its Frobenius norm as 2.02. This data set is a $3\times 6$ matrix, where $d=3$ and $n=6$. There are 7 zero cells in the matrix, showing a medium sparse structure. The mean of these 6 observations (columns) is $(0.27, 0.38, 0.33)$. We use PCA, SVD, NMF and the new NNCA approaches to estimate rank 1 and rank 2 approximation matrices. These approximation matrices by different dimension reduction methods are provided in Table \ref{appmattable}. Note that for the PCA method, the rank 1 matrix is the mean plus the first PC projections, and the rank 2 matrix is the mean plus the first two PC projections. For the SVD and NNCA methods, the rank $k$ matrices ($k=1, 2$) are the cumulative approximation matrices. We observed that the NMF method usually does not provide a unique approximation matrix for any specific rank. Thus, we repeat the NMF calculation 100 times, and summarize the best approximation matrices in terms of smallest Frobenius norm of the corresponding residual.

From Table \ref{appmattable}, we can observe that, both rank 1 and rank 2 approximation matrices from the PCA method contain 3 negative observations, highlighted using a bold font in the table. The rank 2 approximation matrix of the SVD method also contains 3 negative values. Note that no cell in these approximating matrices (of PCA and SVD) is zero, i.e., the approximating matrices are not sparse, which is different from the original input matrix.  Because NMF may generate different approximations (at the same rank) when it is re-applied to the same data set, we report the best approximating matrix of these 100 approximation matrices (last row in Table \ref{appmattable}), which has the smallest Frobenius norm among all the residual matrices. It shows that these approximation matrices are nonnegative. The rank 2 approximating matrices is sparse. However, the subspaces that these two matrices lie in are not nested. The principal angle (see a definition in Section 5) between them is 4.96 degree, showing an important drawback of current NMF algorithms. The NNCA method has both approximation matrices within the nonnegative cone, and their corresponding subspaces are nested. In addition, the rank 2 NNCA approximation matrix shows sparse structure in it as well. All these approximations are used in the visualization of Figure \ref{appmatfig} in Section \ref{sec:intro}.


\subsection{Repeated Simulation Examples \label{subset:repeatrun}}
We simulate 100 examples by using the same simulation settings as in subsection \ref{subsec:onerun}, but with different random numbers for those non-zero entries. Summary measures for all four methods will be provided in this subsection.  Note that for NMF, because of the non-uniqueness of the approximation matrix, we considered100 random restarts for each simulated data set, and choose the best approximating matrix among these.

\begin{table}[ht]
\caption{\small The average number of projections per data set that leave the nonnegative cone, out of $n = 6$.  This highlights serious unsuitability of conventional methods in this situation.} \label{pcasvddrawbacktable}
\begin{center}
\begin{tabular}{l|c|c} \hline\hline
Method & Rank 1 & Rank 2 \\ \hline
PCA & 2.7700 (0.9625) & 3.3200 (0.8025) \\
SVD & 0 (0) & 3.4300 (0.7818) \\ \hline\hline
\end{tabular}
\end{center}
\end{table}

For the PCA/SVD methods, we will report the number of projections that are outside the nonnegative cone. Table \ref{pcasvddrawbacktable} provides a summary of these. It shows that on average, 2.77 observations will be out of the cone for rank 1 PCA approximations, while 3.32 observations will be out of the cone for rank 2 PCA approximations. This indicates that these approximations suffer severely in terms of interpretability. The rank 1 SVD approximation does not have any observations out of the nonnegative cone, which is theoretically supported by the Perron-Frobenius theorem (See e.g. \cite{berman1979nonnegative}). However, the rank 2 SVD approximation has on average 3.43 observations out of the nonnegative cone, i.e. suffers from similar interpretation challenges.

Next we investigate the degree to which the NMF approximations are nested by studying the principal angles between the  subspaces generated by the best rank 1 and 2 approximating matrices.  The principal angle between subspaces  ~\citep{golub1996matrix} are defined as the following. Specifically, let ${\mathcal A} =\mathrm{span}(A_1, \cdots, A_k)$
denote the linear subspace spanned by the columns of $\bA$, and ${\mathcal B}=\mathrm{span}(B_1, \cdots, B_k)$ denote the linear subspace spanned by the columns of $\bB$ respectively. The principal angle between $\cA$ and $\cB$ can be computed as $\mathrm{cos}^{-1}(\rho) \times 180/\pi$, where $\rho$ is the minimum
eigenvalue of the matrix $Q^T_{\bf A} Q_{\bf B}$ where $Q_{\bf A}$ and $Q_{\bf B}$ are orthogonal basis
matrices obtained by the QR decomposition of the matrices ${\bf A}$ and ${\bf B}$, respectively. The result is summarized in Table \ref{avestocvartable}, which shows that the minimal angle between these subspaces is 0.05, i.e, the corresponding subspaces are never nested within each other in any of our simulated realizations. The largest angle between them is around the surprisingly large value of 17 degrees.  

In addition to the above measure, we also investigate the sparsity of the NMF results. The sparsity measure we used in this section is the number of projections that has zero in any of their entries. Note that usually the rank 1 NMF approximation is the same or close to the rank 1 SVD approximation. Thus, in this section, only the sparsity of the rank 2 NMF approximation is reported. The NMF rank 2 approximating matrices on average have 2.91 projections that are sparse, showing NMF has a high chance to get sparse low rank approximations. We also observed that for some realizations,, the approximating matrix was not sparse at all. 

\begin{table}[ht]
\caption{\small The summary of the NMF and NNCA results. For the NMF method, the principal angle between rank 1 and 2 approximating matrices shows that the two corresponding subspaces are never nested with each other. The sparsity measure shows that NMF rank 2 matrices are highly likely to be sparse. The NNCA method can achieve more sparse approximations, compared to the NMF method. } \label{avestocvartable}
\begin{center}
\begin{tabular}{l|l|cccc} \hline\hline
Method & Measure & Min & Max & Mean & Median \\  \hline
NMF & Angle between ranks  (in degree)&  0.05 & 16.64 & 3.80 & 3.55 \\ \cline{2-6}
& Number of sparse projections  & 0 & 4 & 2.91 & 3 \\ \hline \hline
NNCA & Number of sparse projections & 2 & 5 & 3.43 & 3 \\ \hline \hline
\end{tabular}
\end{center}
\end{table}

The NNCA method will generate nested approximating matrices at different ranks. And thus, the principal angle between them will always be zero. The sparsity of the rank 2 approximating matrices is also recorded for these 100 simulated realizations, which is summarized in the last row of Table \ref{avestocvartable}. It shows that the NNCA method generated more sparse projections than NMF. The smallest number of sparse projections is 2. The average number of sparse projections is 3.43, which is larger than that of NMF.  This suggests that NNCA preserves the nice property of NMF: generating sparse approximation.

\subsection{Nonnested structure of NMF \label{subsec:nonnested}}

Another drawback of NMF is that it is not nested.  In this subsection, we investigate this issue more deeply, using a set of simulations to investigate the angle between the rank 1 and rank 2 NMF approximations.  Note that for any nested method, such as PCA and NNCA, this angle is 0, because the rank 1 approximation is contained in the rank 2 approximation. We study a range of different sample sizes $n$, and dimensions $d$. Under each setting, we simulate $n$ $d\times1$ random vectors, where the elements are iid $U[0, 1]$. For each vector, we normalize it to have unit $L_2$ norm, so that every data point lies on the unit sphere. Then we apply the NMF algorithm to find a set of rank 1 and 2 approximations. The angle between these approximations is recorded. We repeat this simulation 100 times for each combination of $n$ and $d$. The summary of such angles are reported in Table \ref{anglestable}.

\begin{table}[ht]
\caption{Summary of the angles (in degrees) between the rank 1 and 2 NMF approximations. Each cell provides the mean angle and the maximum angle of the 100 repeated simulations. This shows that when the dimension increases, the angle between NMF rank 1 and 2 approximations tends to increase.} \label{anglestable}
\begin{center}
\begin{tabular}{l|l|c|c|c|c}\hline\hline
$n$ & \multicolumn{5}{c}{$d$} \\ \cline{2-6}
& &10&100&1000&10000\\ \hline
& Mean&0.33369&1.7112&2.8483&4.4283\\ 
10 & Max&1.8274&2.4654&3.1778&4.6434\\ \hline
&Mean&0.11527&0.94675&3.3778&3.8638 \\ 
100&Max&0.73783&2.168&4.0392&4.8859 \\ \hline
\end{tabular}
\end{center}
\end{table}

Table \ref{anglestable} shows two summaries of such angles: the average angle of the 100 repeated calculations and the maximum of these 100 angles. An important point is that no angle in any of these simulations is exact zero, i.e., the NMF rank 1 and 2 subspaces were never nested in any of our simulated realization. In addition, we also observed that both the average angles and the maximum angles tend to increase as the dimension increases. 

\section{Conclusion and Discussion \label{sec:discuss}}
In this paper, we proposed a novel backwards PCA based NNCA approach. It naturally generates a nested structure in the sequence of approximating cones, giving a major improvement over conventional NMF. In addition, all the approximation matrices are in the nonnegative cone, and thus have better interpretability over traditional PCA/SVD methods.

Note that the nonnegative approximation from rank $k$ to $k-1$ may not be optimal in the sense of the smallest residual. The rank minimization problem is a challenging optimization problem. Advances in optimization may lead to improvements of solving the problem in \eqref{nncaproblem}. We are investigate other potentially better algorithms, and will report them in future papers.

The interpretation of the NNCA sequence $\{\bA_k\}$ is another important challenge for data analysis. An interesting open problem is the adaption of visual devices such as scatter plots of the projection scores, and loadings plots. Note that the nested structure between $\bA_k$'s provides a multi-resolution view of $\bX$, which will form the basis of exploratory methods built on NNCA. We are working visualization tools in this direction as well.

\section*{Acknowledgment}
The first author specially thanks Statistical and Applied Mathematical Sciences Institute (SAMSI) for the kind invitation to participate in two related programs: the Analysis of Object Data and the Massive Data programs. The research of the second author was partially supported by National Science Foundation grant DMS-1109099. This research of the third author was partially supported by National Science Foundation Grant DMS-0854908, and by the SAMSI program on Analysis of Object Data, 2010-2011.

\bibliography{nnca-csda}

\begin{appendix}

\section{Proof of Proposition 1}
\begin{proof}
Since $\bB$ is of rank $k+1$, it can be factorized by SVD as $\bB=\bU_B\bS_B\bV_B^T$. Let $s_1 \geq s_2 \geq \cdots \geq s_k > s_{k+1}>0$ be the singular values in $\bS_B$. Let $u^b_i$ be the $i$th column of $\bU_B$, and $v^b_i$ be the $i$th column of $\bV_B$. From Theorem I in \cite{eckart1936approximation}, we know that $\sum_{i=1}^k s_i u^b_i(v^b_i)^T$ provides the best rank $k$ approximation. i.e.,
\[
\sum_{i=1}^k s_i u^b_i (v^b_i)^T = \mathop{\mathrm{argmin}}_{\mathrm{rank(\bA)=k}} \|\bB-\bA\|_F^2
\]
Define the rank $k$ approximating submatrices $\bU_s=[u^b_1, \cdots, u^b_k]$, i.e., the matrix whose columns are $u^b_1, \cdots, u^b_k$. Let $\bV_s=[s_1v^b_1, \cdots, s_k v^b_k]$. 
Next, we prove that $\mathrm{span}(\bU_s)$ has the smallest square distance from $\bB$, among all rank $k$ subspaces. To this end, let $b_1,\cdots, b_n$ be columns of $\bB$, and let $\mathcal{S}$ be a rank $k$ subspace. Let $y_1,\cdots,y_n$ be $d$-dimensional vectors in $\mathcal{S}$ such that $d(b_i, \mathcal{S})=d(b_i, y_i)$. We have $d^2(\bB,\mathcal{S})=\sum_{i=1}^n d^2(b_i,y_i) = \|\bB-\bA\|_F^2$, where $\bA$ is the matrix whose columns are $y_1,\cdots,y_n$. Since the rank of $\bA$ is no more than $k$, we know that $\|\bB-\bA\|_F^2 \ge \|\bB-\bU_s \bV_s^T\|_F^2=d^2(\bB,\mathrm{span}(\bU_s))$. The fact that $\bU_s \bV_s^T$ is the unique best rank $k$ approximation of $\bB$ implies that $\mathrm{span}(\bU_s)$ is the unique rank $k$ subspace that has the smallest square distance from $\bB$.
\end{proof}
\section{Proof of Proposition 2}
\begin{proof}
Since $s_k>s_{k+1}$, the Theorem I in \cite{eckart1936approximation} implies that $u^b_k$ is uniquely defined (up to a sign change). Without loss of generality, assume that all the first $k$ singular values $s_i$ are different. Then $u^b_i$ are uniquely defined as well (up to the same sign change). It remains to prove that $\bV$ is uniquely defined.

The optimization problem is to minimize $\|\bB-\bU_s\bV^T\|_F^2$ subject to $\bU_s\bV^T\geq 0$. Note that $\|\bB-\bU_s\bV\|_F^2$ is a quadratic function of $\bV$. Moreover, since $U_s$ is of rank $k$, the matrix $U_s^T U_s$ is positive definite, so $\|\bB-\bU_s\bV\|_F^2$ is a strictly convex function, see, e.g., \cite{mag.neu:mdc}.
 The constraint $-\bU_s\bV^T \leq 0$ is a standard linear inequality constraint on $\bV$, and $\bV={\bf 0}_{n\times k}$ obviously satisfies this constraint. Thus, by convexity theory, this optimization problem has a unique solution. 

\end{proof}
\end{appendix}
\end{document}